\documentclass[runningheads]{llncs}
\usepackage[T1]{fontenc}
\usepackage{amssymb}

\usepackage{todonotes}
\usepackage{hyperref}
\usepackage{cite}
\usepackage{comment}
\usepackage{algorithm}
\usepackage{algpseudocode}
\usepackage[beginComment=//~,beginLComment=//~,endLComment=~,]{algpseudocodex}

\usepackage{graphicx}    
\usepackage{pdflscape}   

\newcommand{\curly}{\mathrel{\leadsto}}

\newcommand{\ceep}{{\sc ceep} }
\newcommand{\cevp}{{\sc cevp} }

\newtheorem{Observation}{Observation}

\begin{document}

\title{Popping Bubbles in Pangenome Graphs}

\author{Njagi Mwaniki\inst{1}\orcidID{0000-0002-4858-2375} \and
Erik Garrison\inst{2}\orcidID{0000-0003-3821-631X} \and
Nadia Pisanti\inst{1}\orcidID{0000-0003-3915-7665} 
}

\authorrunning{Mwaniki, Garrison, Pisanti} 

%
\institute{Department of Computer Science, University of Pisa, Italy
\email{njagi.mwaniki@di.unipi.it} \&
\email{nadia.pisanti@unipi.it} \and
University of Tennessee Health Science Center, USA\\
\email{egarris5@uthsc.edu}}




\maketitle              
\begin{abstract}
In this paper we introduce \emph{flubbles}, a new definition of "bubbles" that correspond to variants in a (pan)genome graph $G$. We then show a characterization for flubbles in terms of equivalence classes regarding cycles in an intermediate data structure we built from the spanning tree of the $G$, which leads us to a linear time and space solution for finding all flubbles. Furthermore, we show how a related characterization allows us to efficiently detect also what we define \emph{hairpin inversions}: a cycle preceded and followed by the same path in the graph; being the latter necessarily traversed both ways, this structure corresponds to inversions. 
Finally, Inspired by the concept of Program Structure Tree introduced fifty years ago to represent the hierarchy of the control structure of a program, we define a tree representing the structure of $G$ in terms of flubbles, the \emph{flubble tree}, which we also find in linear time. The hierarchy of variants introduced by the flubble tree paves the way to new investigations of (pan)genomes structures and their decomposition for practical analyses.
We have implemented our methods into a prototype tool named {\rm povu} which we tested on human and yeast data. We show that {\rm povu} can find the flubbles, and also output the flubble tree, while being as fast (or faster than) well established tools that find bubbles such as {\rm vg} and {\rm BubbleGun}. Moreover, we show how, within the same time, {\rm povu} can find hairpin inversions that, to the best of our knowledge, no other tool are able to find. Our tool is freely available at 
{\rm https://github.com/urbanslug/povu/} under the MIT License.

\end{abstract}

\newpage

\section{Introduction}
\label{sec:intro}

A \emph{pangenome} was defined in~\cite{consortium_computational_2018} as \emph{any collection of genomic sequences to be analyzed jointly or to be used as a reference} in 2018. 
Since then, computational pangenomics faces the algorithmic issues of the paradigm shift from linear reference genomes, to more complex graph-like structures required for pangenomes, with the suggestion of new pangenome representations and new tools to investigate them. 
On the representation side, many graph-like structures that can be directed or not directed, labelled or not labelled, coloured or not coloured, and acyclic or not, have been suggested. 
Furthermore, for each possible representation  efficient tools to solve basic toolkit problems such as mapping a read therein, as well as more general tasks such as detecting variants, have also been suggested. 

The challenge of a trade off emerges: on one hand the choice of complex structures that allow to represent more kinds of genomic variants at a cost of being hard to analyze (and to visualize), and on the other hand that of simpler linearized versions that fail to express structural variants but exhibit algorithmic advantage supporting fast search therein.

In any of the possible pangenome graph representation, a genomic variant is a so-called \emph{bubble}: a set of distinct possible paths (the variants) that share the same contexts at their sides. The notion of bubble~\cite{bubble02} has been extensively investigated in the literature even before computational pangenomics was conceived. Even for bubbles we have witnessed an explosion of possible definitions and of possible algorithms to detect them: superbubbles~\cite{onodera_detecting_2013}, ultrabubbles~\cite{zerbino_velvet_2008}, mouths~\cite{mouth-spire10}, snarl~\cite{paten_superbubbles_2018}, if not more. The algorithms that have been suggested  to find these structures are either quadratic or bound to work on acyclic graphs.\\

In this paper we provide an overview of some of the main notions of both (pan)genome graphs and bubbles. Then we choose \emph{variation graphs}, that are (equivalent to one of) the most general pangenome graph representations, and we define therein a new notion of bubble, which we name \emph{flubble}.
We thus describe an intermediate data structure that we build from the spanning tree of the variation graph, and we give a characterization for flubbles in terms of equivalence classes regarding cycles such data structure, which leads us to a linear time and space solution for finding all flubbles.
Furthermore, we show how a related characterization allows us to efficiently detect also what we defined \emph{hairpin inversions}: a cycle preceded and followed by the same path in the graph; being the latter necessarily traversed both ways, and hence one forward and one reverse (one before and one after the cycle), this structure corresponds to inversions. 
Finally, Inspired by the concept of Program Structure Tree introduced in \cite{compiler94} with the purpose of representing the hierarchy of the control structure of a program, we define a tree representing the structure of a variation graph in terms of flubbles, the \emph{flubble tree}, which we also find in linear time. In our tree, nodes are flubbles, whereas edges are nesting relations among them.
By extending an idea of~\cite{paten_superbubbles_2018} for decomposing graphs in terms of bubbles, we propose the flubble tree to drive a progressive \emph{popping} of flubbles to decompose complex pangenome graphs into simpler structures.
We have implemented our methods into a prototype tool named {\rm povu} which we tested on human and yeast data. We show that {\rm povu} can find the flubbles, and also output the flubble tree, while being as fast (or faster than) well established tools that find bubbles such as {\rm vg} and {\rm BubbleGun}. Moreover, we show how, within the same time, {\rm povu} can find hairpin inversions that, to the best of our knowledge, no other tool is able to find.

\paragraph{\bf Paper organization}
After some preliminary definitions of pangenome graphs (Section~\ref{sec:graphs}) and of bubbles (Section~\ref{sec:bubbles}), we introduce our results (Section~\ref{sec:povu}) with the characterization of hairpin inversions  (Section~\ref{ssec:hair}), and of flubbles with their flubble tree (Section~\ref{ssec:tree}), for which we also suggest some possible applications (Section~\ref{sec:decompose}). We conclude with experimental results  over the efficiency and accuracy of our tool (Section~\ref{sec:expe}).

\section{Graphs in (pan)genomics}
\label{sec:graphs}

A Graph $G=(V,E)$ consists of a set $V$ of vertices and a set $E$ of edges. An edge $e\in E$ is a pair $(i,j)$ 
where $i,j\in V$. 
When such pair is ordered we say that $G$ is \emph{directed}, and it is \emph{undirected} otherwise. 
If $i=j$ then the edge $(i,j)$ is called a \emph{self-loop}.
A \emph{walk} in a graph is a finite sequence $v_1, v_2, \cdots, v_k$ of vertices of $V$ such that $(v_i,v_{i+1})\in E$ for each $1\leq i \leq k-1$. We will also denote with $v_1 \curly v_k$ a walk from $v_1$ to $v_k$.

In genomics, graphs are often used to represent genomic sequences by means of a labeling of vertices or edges. In this paper we will assume a vertex labeling as follows: $l(v)$ be a function that associates a string (which we will also name \emph{sequence}) in the alphabet $\{A,C,G,T\}$ to a vertex $v\in V$. 
The labeling extends to walks by concatenating the labels of the vertices: $l(v_1, v_2, \dots, v_k)=l(v_1)\cdot l(v_2)\cdot ... \cdot l(v_k)$.

A \emph{Bidirected Graph $G_d=(V_d,E_d)$} \cite{KS95,paten_superbubbles_2018} is a graph in which each end of an edge in $E_d$ is either a head or a tail, that is, every edges has an independent orientation indicating if the endpoint is incident with the head or the tail of a given vertex; each edge in $E_d$ is a pair of elements of the set $V_d \times \{head,tail\}$.
We say that $(x,head)$ and $(x,tail)$ are \emph{opposite sides} of $x\in V_d$. We remark that, even if a bidirected graph is not a multigraph, two vertices $u$ and $v$ therein can be connected by two distinct edges when the edges differ because of their $\{head,tail\}$ field.\\
A \emph{Sequence Graph} \cite{PZHH14,novak_genome_2017,paten_genome_2017} is a bidirected graphs in which each vertex $v$ is labelled with a non empty string $l(v)$. In a sequence graph, a walk $v_1,v_2,\dots , v_k$ are considered \textit{valid} if for each $v_i$, if the edge connecting $v_{i-1}$ and $v_i$ enters $v_i,head$ (resp. $v_i,tail$) then the edge connecting $v_{i}$ and $v_{i+1}$ must exit $v_i,tail$ (resp. $v_i,head$). That is, a walk must exit each vertex through the opposite side in which it entered, else the walk is \textit{invalid}.

In a \emph{Biedged Graph $G_e=(V_e,E_e)$} \cite{paten_superbubbles_2018} $E_e$ contains two types of edges: \emph{black edges} and \emph{grey edges}, and such that each vertex in $V_e$ is incident with at most one black edge. 
The set $V_e$ contains vertices $(x,tail)$ and $(x,head)$ for as many $x$ as $V_e/2$, and black edges connect $(x,tail)$ and $(x,head)$ for each $x$.
In the biedged graph, a walk is considered \textit{valid} if it alternates between black and grey edges, and considered \textit{invalid} otherwise.

As shown in \cite{paten_superbubbles_2018}, the biedged graph can be seen as an alternative representation of a bidirected graph where each vertex $x\in V_d$ is split in two in $V_e$: $(x,tail)$ and $(x,head)$, and black edges of $E_e$ connect these two opposite sides of $x$. In a biedged graph, labels that associate strings to walks are indeed defined on black edges.
Conversely, to go from a biedged graph to a bidirected graph one would only need to collapse each pair of nodes $(x,tail)$ and $(x,head)$ of $V_e$ connected by a black edge into a single node $x$ of $V_d$ with head and tail sides. 
Therefore, the bidirected and biedged representations are equivalent, and the choice to use one is based on the application.
Bidirected and biedged graphs have been introduced with the aim of allowing both forward and reverse traversal of nodes according to whether we are considering the forward or reverse strand of DNA (in bidirected graph labels are associated to vertices, whereas in biedged graphs they are associated to black edges). 

In this paper we refer to Variation Graphs  \cite{vg}, as biedged versions of Sequence Graphs.




A \emph{cycle} in a graph is a path $v_1 \curly v_k$ in which $v_1=v_k$, and it is a \emph{simple cycle} if no other vertex is repeated. 
A Cactus Graph \cite{harary_number_1953,DBLP:conf/recomb/PatenDEJMSH10} is a multigraph in which 
each edge is part of at most one simple cycle, and hence any two simple cycles intersect at most one vertex \cite{paten_superbubbles_2018}.

A spanning tree of a graph $G=(V,E)$ is a subgraph of $G$ that contains every vertex in $V$ and is a tree. A depth-first search (DFS) spanning tree $T_s$ is a spanning tree generated from the depth-first traversal of a graph starting from $s\in V$.

\begin{definition}[Edge Cycle Equivalence and Node Cycle Equivalence]
\label{def:Edge and Node Cycle Equivalence}
Any two edges $a,b\in E$ in a connected component of a graph are \emph{edge cycle equivalent} if and only if every cycle containing $a$ also contains $b$, and vice versa. In this case we will say that $a$ and $b$ are a \ceep (Cycle Equivalent Edge Pair).
Similarly, any two nodes $u,v\in V$ in a connected component of a graph are said to be \emph{node cycle equivalent} if and only if every cycle that contains $u$ also contains $v$, and vice versa. In this case we will say that $u$ and $v$ are a \cevp (Cycle Equivalent Vertex Pair).
\end{definition}
We remark that there is a one-to-one correspondance betwen CEEPs of black edges in a biedged graph and CEVPs of opposite vertices in the equivalent bidirected graph.

\begin{definition}[k-edge-disconnectable and k-vertex-disconnectable]
\label{def:k-edge-disc}
A proper subgraph $G_s$ of a graph $G$ is said to be \emph{k-edge-disconnectable} if it can be disconnected from $G$ by removing $k$ edges, and $G_s$ cannot be disconnected from $G$ by removing less than $k$ edges.
Similarly, a proper subgraph $G_s$ of a graph $G$ is said to be \emph{k-vertex-disconnectable} if it can be disconnected from $G$ by removing $k$ vertices, and $G_s$ cannot be disconnected from $G$ by removing less than $k$ vertices.
\end{definition}



\section{Bubbles}
\label{sec:bubbles}

The concept of bubble in a graph was introduced in \cite{bubble02} as 
"a region in which multiple paths exist" in a directed graph. In \cite{paten_superbubbles_2018} a bubble is defined as a pair of paths that connect a common source node $s$ to a common target node $t$ that are otherwise disjoint (that is they do not share any node other than $s$ and $t$).
Bubbles were also defined as \emph{mouths} for the special case of SNPs in de Bruijn Graphs of order $k$ as two distinct paths of length $k$ from a common source vertex to a common target vertex \cite{mouth-spire10}.
The concept was later extended to allow multiple paths from $s$ to $t$, and/or the case $s=t$, and possibly nested structures \cite{bubble-spire-12}. An important step towards efficiency came in \cite{onodera_detecting_2013} with the definition of \emph{superbubbles} as their properties restrict their number to $\mathcal{O}(n)$ where $n=|V|$. Superbubbles are defined as follows in a directed graph where we say that a node $v$ is \emph{reachable} from a node $u$ if there is a walk from $u$ to $v$.
A \emph{superbubble} \cite{onodera_detecting_2013} is as any pair of distinct vertices $<\!s,t\!>$ in a directed graph which satisfy the following properties:
\begin{itemize}
    \item \textit{reachable}: $t$ is reachable from $s$,
    \item \textit{matches}: the set of vertices $U$ reachable from $s$ without passing through $t$ is equal to the set of vertices from which $t$ is reachable without passing by $s$, 
    \item \textit{acyclic}: the subgraph induced by $U$ is acyclic, and
    \item \textit{minimal}: no vertex in $U$, other than $t$, forms a pair with $s$ that satisfies the conditions above.
\end{itemize}

The definition of superbubbles was then extended to biedged graphs in \cite{paten_superbubbles_2018}: (i) a \emph{snarl} is defined as the minimal subgraph of a biedged graph with the \emph{2BEC} property (that is, being \emph{2-black-edges connected}), which means that it cannot be disconnected by removing less than $2$ black edges (assuming grey edges cannot be removed), and (ii) an \emph{ultrabubble} is tip-free acyclic snarl (a \emph{tip} in a biedged graph is any node that is not incident with a grey edge \cite{zerbino_velvet_2008}).\\

The motivation for all above mentioned notions of bubble that have been suggested, as well as for the many algorithms and tools that have been devised to enumerate them, is that they represent genomic variants as: (i) they correspond to subgraphs that are identified by a pair of nodes that correspond to the common left and right context of the variant, and (ii) the possible traversals of this subgraph spell distinct sequences that indeed represent the variants.

\section{The Algorithm {\rm povu}}
\label{sec:povu}

Our algorithm {\rm povu} takes in input a Variation Graph $G=(V,E)$. We assume without loss of generality that $G$ is compact, that is there are no linear chains containing more than one black edge: should this be the case, they could be collapsed into a single black edge obtaining an equivalent graph.

As a first step, should $G$ not be connected, {\rm povu} decomposes it into connected components to be processed separately. Therefore, from now on we (can) assume that the graph $G$ is connected.

As a second step, {\rm povu} detects \emph{tips} and creates as many dummy (grey) edges as the number of detected tips to connect each one of them to a suitably created dummy node $s$. Should $G$ not have tips, let $s$ be any node of $G$.

Third, a (pre-order) DFS spanning tree $T_s$ is computed for $G$ using $s$ as a root.

Before describing our methods to detect hairpins and compute the flubble tree, we make the following observations and definitions:
\begin{itemize}
    \item The spanning tree $T_s$ introduces a partial order $\prec$ over the vertices of $G$: we say that $v_1 \prec v_2$ if $v_1$ and $v_2$ belong to the same root-leaf path $T_s$ and $v_1$ is closer to $s$.
    \item We remark that any edge $(u,v)\!\in\! E\!\setminus\! T_s$ is such that $u\!\prec\! v$ or $v\! \prec\! u$. We name these edges \emph{back-edges}, and we denote with $T'_s$ the tree $T_s$ augmented with back-edges (that is, $T'_S$ is basically $G$ with the partial order $\prec$).
    \item A \emph{bracket} of a tree edge $e \in T_s$ is a back-edge of $T'_s$ connecting a descendant of $e$ to an ancestor of $e$. 
    \item Two edges $e_1,e_2 \in T_s$ are cycle equivalent  in $T'_s$ if and only if they contain the same sets of brackets.
\end{itemize}

\subsection{Detecting Hairpins (Inversions)}
\label{ssec:hair}

\begin{figure}
\centering
\includegraphics[width=0.8\textwidth]{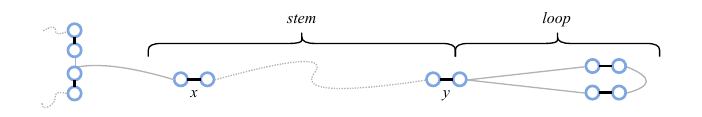}
\caption{An example of hairpin in a biedged graph with boundaries $x$ and $y$.}
\label{fig:hairpin}
\end{figure}

Inspired by a similar shape in RNA secondary structure, we define a \emph{hairpin} in a biedged connected graph as a 1-blackedge-disconnectable subgraph containing a 1-blackedge-disconnectable cycle. A hairpin comprises two parts (see Figure~\ref{fig:hairpin}):
\begin{itemize}
    \item a \textit{loop}, a 1-blackedge-disconnectable simple cycle. 
    \item a \textit{stem}, the cycle free 2-blackedge-disconnectable subgraph connected to the loop, or the single edge connecting the loop to the rest of the graph;
\end{itemize}

Structural variation plays a major role in genetic diversity within the human genome and is believed to account for the majority of differences between individuals. Further, it is known to have profound consequences in evolution and human disease  \cite{feuk_structural_2006, sudmant_integrated_2015, collins_structural_2020}. 
From a biological standpoint, inversions can be expressed as hairpins in a variation graph. From a theoretical standpoint, hairpin inversions (depending on  the size of the hairpin stem) are possibly complimented palindromes~\cite{gusfield_algorithms_1997}. The hairpin detection problem is related to (but not exactly) the problems of finding articulation points or bridges\footnote{Indeed, nodes inside the stem of a hairpin are \emph{articulation points}, and the edges therein are \emph{bridges} for the biedged graph $G$~\cite{cormen_introduction_2009}.} in a graph~\cite{even_graph_2012, tarjan_depth-first_1972}. Nevertheless, for hairpin detection we will employ $T'_s$ making use of the following observation.

\begin{Observation}
\label{trm:hairpin stem lacks backedge} 
A hairpin in $G$ is obviously a hairpin in $T'_s$. Moreover, if $G$ contains a hairpin, then at least one edge of $T_s$ will lack a bracket in $T'_s$.
\end{Observation}

An example of hairpin with boundaries $<14,15>$ is in the biedged graph of Fig.~\ref{fig:example}(a) as well as in its $T'_s$ shown in  Fig.~\ref{fig:example}(b).

\subsection{The Flubble Tree}
\label{ssec:tree}

Rather than just enumerating the flubbles, our algorithm implicitly lists them as nodes of a tree $T_F$, the \emph{flubble tree}, which will additionally describe the nesting structure of the flubbles of $G$. 

Flubbles are found by assigning cycle equivalence classes\footnote{A cycle equivalence class is a set where all elements are pairwise cycle equivalent. This is indeed an equivalence class.} to edges in $T'_s$, and thus detecting a flubble per each equivalence class. 
We will identify a flubble of $G$ with a pair $<\!e_1,e_2\!>$ (named \emph{boundaries of the flubble}) of distinct black edges of $T_s$ (and hence of $G$) that are cycle equivalent in $T'_s$. Indeed, being $T_s$ a tree, there are no cycles unless we consider its augmented version, but only edges of $T_s$ can be boundaries of a flubble. 
The flubble boundaries are detected in linear time by traversing $T_s$ in reverse DFS order, as the two black edges that open and close the cycle equivalence class. 

This requirements implicitly ensures the following properties we desire for a flubble:
\begin{itemize}
\item Being $T_s$ a DFS spanning tree ensures that the back-edges will not overlap eachother, and hence edges will lie in cycle equivalence classes.   
\item Being $T_s$ a DFS spanning tree, and being $G$ compact, these two edge belong to the same root-leaf path in $T_s$, and there are multiple paths connecting these two edges.
\item The pair $<\!e_1,e_2\!>$ make $G$ 2-blackedges-disconnettable, and hence each flubble is a \ceep. 
\end{itemize}

More importantly, since we can prove that the number of back-edges is upper bounded by (and actually strictly less than) the number of edges in $T_s$, and since there are at most as many flubbles as back-edges (possibly +1), then we can prove the following results.

\begin{theorem}
In a biedged graph $G=(V,E)$, there are at most a $|E|$ flubbles.
\end{theorem}

\begin{corollary}

The flubble tree can be computed in linear time in the size of $G$.
\end{corollary}

\begin{figure}
\centering
\includegraphics[width=1.0\textwidth]{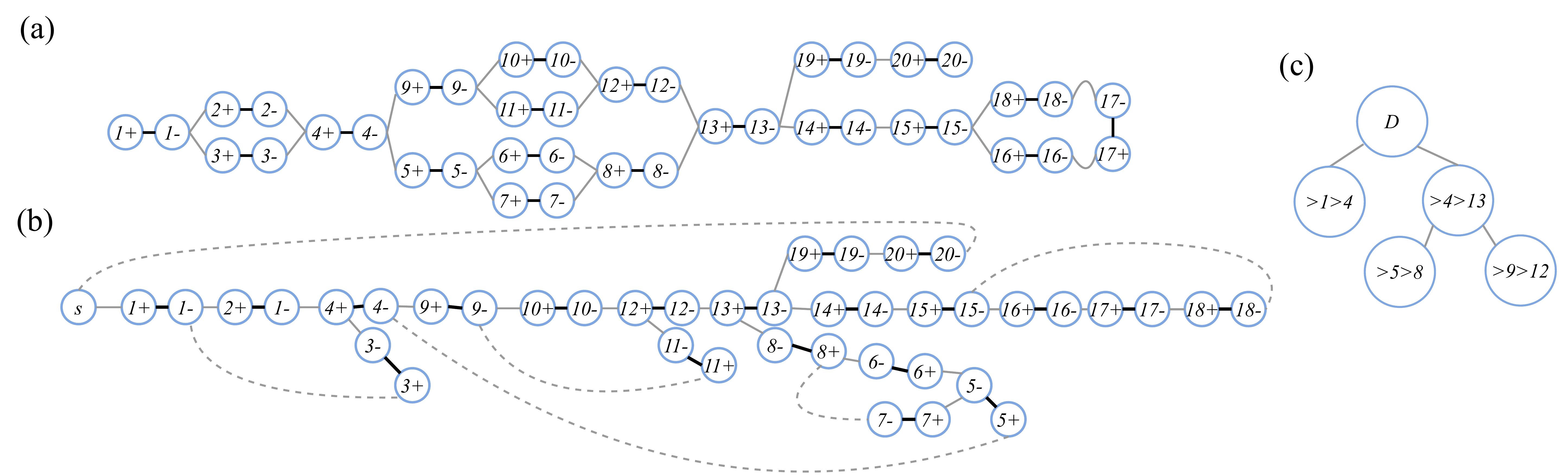}
\caption{A biedged graph (a), its augmented spanning tree (b), and its flubble tree (c). An hairpin with boundaries $<14,15>$ would also be reported.}
\label{fig:example}
\end{figure}

An example of flubble tree is shown in Fig.~\ref{fig:example}(c) for the biedged graph of Fig.~\ref{fig:example}(a). The augmented tree $T'_s$ for the same graph is shown in Fig.~\ref{fig:example}(b), where we still have the hairpin (with boundaries $<14,15>$) that will be reported as such, and the tip of black edges 19 and 20 that will be disregarded. Indeed, the flubble (tree) detection will actually take place on a suitably modified version of such $T'_s$ that will add some further virtual backedges that will allow to distinguish hairpins and tips from flubbles.

\section{Decomposing Pangenome Graphs via the Flubble Tree}
\label{sec:decompose}

The topological properties of $T_F$ will necessarily reflect the (pan)genome that $G$ represents. The leaves of $T_F$ correspond to \emph{minimal} flubbles, that is, those that do not contain any other flubble: these can be SNPs or INDELs or in general a list of alternative strings. Should there be a chain of them, these will be siblings leaves whose parent is a flubble that contains them. Hence, the wider the tree is, the more consecutive flubbles there are. Furthermore, the deeper $T_F$ is, and the more nested bubbles there are. An interesting use of the flubble tree could be to use it as representing structural features of a (pan)genome.

Another possible use of the flubble tree could be to drive a sort of progressive decomposition of a (pan)genome graph by \emph{popping} the flubbles via a post-order visit of $T_F$.
Indeed, computational pangenomics~\cite{consortium_computational_2018} is facing the challenge of finding the trade off between pangenomes representations that allow to highlight complex variants~\cite{vg,vghaplo,novak_genome_2017,alfapang24} but are computationally difficult to index or to search~\cite{DBLP:journals/talg/EquiMTG23}, and others - such as (E)D strings~\cite{fi20,cpm17} or founder graphs~\cite{DBLP:journals/tcs/RizzoENM24} - that have limited expressive power but allow fast pattern matching~\cite{biostec23,EDS_mapping,sicomp22} (also in the presence of errors~\cite{tcs20,isbra22,biostec23}), and/or indexing~\cite{DBLP:journals/tcs/RizzoENM24}. A post-order visit of the flubble tree could drive a processing of complex graphs that begin by collapsing (linear chains) of minimal flubbles into structures that are easier to handle (for example replacing the labels string in labels with ED strings), and then prune $T_F$ \emph{popping} these leaves (and possibly iterate the process in a progressive manner) turning the graph into a hybrid representation that could combine the benefit of complex graphs and ED strings. 

\section{Experimental Results}
\label{sec:expe}

\begin{figure}[!ht]
\centering
\includegraphics[width=1.0\textwidth]{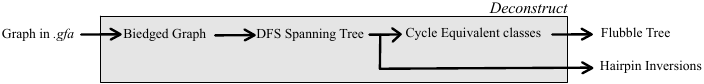}
\caption{Workflow of {\rm povu} \emph{Deconstruct}.}
\label{fig: deconstruct}
\end{figure}

We implemented our method into a tool called {\rm povu} which is freely available at 
{\rm https://github.com/urbanslug/povu/} under the MIT License. The \emph{Deconstruct} command of {\rm povu}, whose workflow is illustrated in Figure~\ref{fig: deconstruct}, takes in input a Variation Graph in {\rm .gfa format} \cite{gfaurl}, reports its hairpins, and generates a flubble forest (that is, a flubble tree per each connected component of the input graph).

We see a {\rm .gfa} graph as a biedged graph with string labels on black edges (Segments lines of the .gfa format, whereas the grey edges are the Link lines). {\rm povu} could as well take in input a Sequence Graph provided this is turned into the equivalent biedged graph. 

\begin{table}[!ht]
\centering
\caption{\label{res: HPRC}Experiments on real data. Time reported in seconds.}
\begin{tabular}{l|l|l|l|l|l}
$Data$ & $|V| $ & $|E|$ & $povu$ & ${\rm BubbleGun}$ & ${\rm vg}$ \\
\hline
Human chromosome 22 & 3,759,736 & 5,224,421 & 110.72 & 90.46 & 226.36  \\
Human chromosome Y & 318,340 & 441,309 & 4.02 & 7.35  &  13.66 \\
\textit{S. cerevisiae} &  603,070 &  815,621 & 6.06 & 14.22 &  7.65 \\
Human chromosome 4 C6 region &  1,748 & 2,366 & 0.0 & 0.01 &  0.02 \\
Human chromosome 6 LPA region & 3,751 & 5,195 & 0.0 & 0.01 &  0.01 \\
\end{tabular}
\end{table}

We compared {\rm povu} to existing tools, specifically {\rm vg}, version 1.57.0, and {\rm BubbleGun}, version 1.1.8. As for the data, we used a yeast dataset of \textit{Saccharomyces cerevisiae} (a complex dataset containing tips and hairpins) and a subset of two graphs from the draft human pangenome \cite{liao_draft_2023} that include the Y chromosome (short but very complex: it was the only one in which the Telomere-to-Telomere consortium could not get a complete assembly of~\cite{nurk_complete_2022}). 
All experiments were run on a laptop Intel® Core™ i7-11800H × 16 with 16.0 GiB RAM. 

The tool {\rm vg} finds snarls, whereas {\rm BubbleGun} detects superbubbles and superbubble chains. Even if these and flubbles are three different bubble definitions, the three tools basically gave the same output with very few differences: snarls can contain tips whereas flubbles don't, but on the other hand 
{\rm vg} seems to miss (some) minimal flubble when they are a chain; {\rm BubbleGun} outputs is quite a small subset of {\rm povu}'s and of {\rm vg}'s unless the {\rm bchain} option is used. As for time, as shown in Table~\ref{res: HPRC}, even if {\rm povu} does the extra work of building the flubble tree, it outperformed {\rm vg} in all datasets,  and only took longer when compared to {\rm BubbleGun} for the human chromosome 22 dataset.


As shown above, the \emph{Deconstruct} command of {\rm povu} can also detect hairpins. To the best of our knowledge, no other tool can detect them in pangenome graphs. We tested this feature on some of the datasets available to us finding few but significant hairpin inversions; results are shown in Table~\ref{tbl:hairpins}.

\begin{table}[!ht]
\caption{\label{tbl:hairpins}Hairpin Inversions}
\begin{tabular}{l|c|l}
Dataset & Count & Pairs of Hairpin Inversion Boundaries  \\
\hline 
Human chrom. 22 & 2 & \footnotesize <3758903, 3758902>,  <2067039, 2066864> \\
Human chrom. Y & 3 & \footnotesize <104681, 104663>, <275736, 276460>, <95708, 95566> \\
\textit{S. cerevisiae} & 1 & \footnotesize <462565, 461860> \\
Human chrom. 4 C6 region &  0 &  \\
Human chrom. 6 LPA region & 0 & \\
\end{tabular}
\end{table}

\begin{figure}[ht]
\centering
\includegraphics[width=0.7\textwidth,height=5cm]{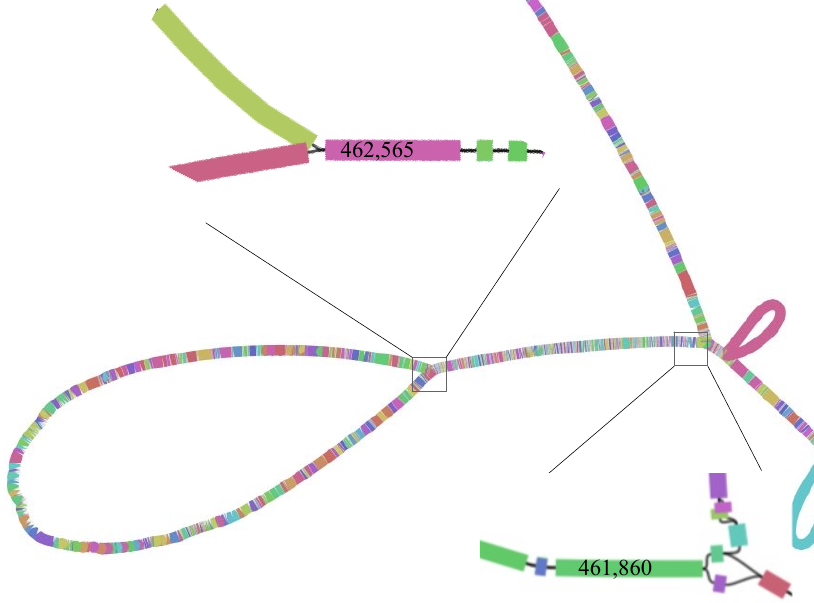}
\caption{A {\sc bandage}~\cite{bandage} visualization of a hairpin $<462565,461860>$ detected by {\rm povu} in \textit{S. cerevisiae} where the two boundaries are highlighted (zoomed in).}
\label{fig: yeast hairpin}
\end{figure}

Figure~\ref{fig: yeast hairpin} shows a visualization of the hairpin boundaries (the edges at the two ends of the stem) we detected for \textit{S. cerevisiae}.

\section*{Acknowledgements}

This work was supported by (i) PANGAIA and ALPACA projects that received funding from the European Union’s Horizon 2020 research and innovation programme under the Marie Skłodowska-Curie grant agreements No. 872539 and 956229, respectively; (ii) NextGeneration EU programme PNRR ECS00000017 Tuscany Health Ecosystem: (iii) the MUR PRIN 2022 YRB97K PINC.\\ 
The results were developed during the visit of EG to NM and NP which was funded by the Vising Fellow program of the University of Pisa in 2023.


\newpage

\bibliography{main}

\end{document}